\documentclass[,final,numberedheadings]
 {aipproc}
\layoutstyle{6x9}
\begin{document}
\title{The connection between mass loss and nucleosynthesis}
\author{Jacco Th.\ van Loon}{address={Lennard-Jones Laboratories, Keele
University, Staffordshire ST5 5BG, United Kingdom}}
\begin{abstract}
I discuss the relationship between mass loss and nucleosynthesis on the
Asymptotic Giant Branch (AGB). Because of thermal pulses and possibly other
mixing processes, products of nucleosynthesis can be brought to the surface of
AGB stars, increasingly so as the star becomes more luminous, cooler, and
unstable against pulsation of its tenuous mantle. As a result, mass loss is at
its most extreme when dredge-up is too. As the high rate of mass loss
truncates AGB evolution, it determines the enrichment of interstellar space
with the AGB nucleosynthesis products. The changing composition of the stellar
atmosphere also affects the mass-loss process, most obviously in the formation
of dust grains --- which play an important r\^ole in driving the wind of AGB
stars.
\end{abstract}
\maketitle

\section{Introduction}

Asymptotic Giant Branch (AGB) stars span a range in masses from $<1$ M$_\odot$
to $\sim8$ M$_\odot$, which makes them important diagnostics of the star
formation history and important players in galactic chemical enrichment from
ages as young as $<100$ Myr up to (nearly) a Hubble time. They shed much,
often most, of their initial mass in the form of dusty winds before leaving a
white dwarf behind.

In this review I will first ask the question why we care about mass loss, with
a bias towards those amongst us who are interested in nucleosynthesis. I then
briefly discuss what determines the mass-loss rate, and comment on ways in
which nucleosynthesis may, or may not, alter the conditions for the mass-loss
mechanism.

\section{Why do we care about mass loss?}

Besides its importance for galactic recycling processes and the build-up of
chemical richesse, mass loss greatly affects the star itself. Although mass
loss may be seen as an outer boundary condition, it does affect the stellar
interior.

As a result of mass loss the convective mantle reduces in mass as well as
density. This enhances the effects from $3^{\rm rd}$ dredge-up, as a given
mass of dredged-up product of nucleosynthesis is going to lead to a larger
enrichment of the mantle (and photosphere) if it is less diluted by
pre-existing material. Massive AGB stars (roughly $M_{\rm init}>4$ M$_\odot$)
experience Hot Bottom Burning (HBB), which converts carbon into nitrogen (and
oxygen), which is the reason why they do not normally become carbon stars.
However, if mass loss depletes the mantle to near extinction, HBB ceases and a
final thermal pulse will almost certainly turn the star into a carbon star.

Once the mantle mass is depleted to the extent that it becomes transparent,
the star leaves the AGB and gradually dies. The higher the rate of mass loss
on the AGB, the earlier this happens and the fewer thermal pulses the AGB
stars will have experienced. This affects the photospheric abundances observed
near the tip of the AGB, as well as the chemical and isotopic yields from AGB
stars. Mass loss will also truncate the growth of the core, and thus affect
the mass distribution of white dwarfs.

Finally, as the convective envelope {\it expands} as a reaction to its
diminished density, the surface gravity and temperature drop, which would
appear to facilitate further mass loss. Stars for which this becomes important
are generally seen to pulsate radially in the fundamental mode, and their
pulsation period increases with increasing radius. If the radius increases to
the extent that the pulsation period approaches the thermal timescale then the
pulsating mantle will continuously adapt its thermal structure as it expands,
and this might detach the mantle to form a planetary nebula (PN).

The observed enrichment of the interstellar medium and subsequent generations
of stars which is due to AGB stars depends on the AGB mass loss. This is true
for elements as well as isotopic ratios; \cite{KarakasEtal06}, for example,
show that the \cite{VassiliadisWood93} mass-loss formalism leads to a 75\%
reduction in the yields of elements like silicon and aluminium compared to
using the Reimers' Law \cite{Reimers75}.

The dust production by AGB stars is also affected by the mass-loss history,
but in a slightly different way \cite{FerrarottiGail06}. We shall later see
how the dust depends on the detailed chemical composition of the star, but it
is already obvious that carbon stars would form very different dust from
other, oxygen-rich AGB stars. Depending on how the mass loss varies in time,
this may be accompanied by more or less dust mass, and more or less of the
dust produced whilst the star is a carbon star (if it ever becomes one). This
may all depend on initial metallicity; \cite{FerrarottiGail06} predict that
metal-poor AGB stars produce predominantly carbonaceous dust also if they are
massive.

The Initial-Final Mass (IFM) relation depends on how quickly the mass loss
truncates the growth of the core on the AGB. Measurements of the IFM relation
in galactic open clusters \cite{Williams07} and central stars of PNe in the
metal-poor Magellanic Clouds \cite{VillaverEtal07} suggest that it varies very
little with metallicity. As a consequence, the total amount of mass shed on
the AGB is quite well known, e.g., $\Delta M=4.1\pm0.1$ M$_\odot$ for $M_{\rm
init}=5$ M$_\odot$. This is because mass loss ($\dot{M}\gg 10^{-7}$ M$_\odot$
yr$^{-1}$) outpaces nuclear burning for long enough ($\sim$0.1--1 Myr),
causing negligible core growth whilst the mantle is shed. This is true for the
solar neighbourhood, with $Z_{\rm init}=$Z$_\odot$, but also in the LMC,
$Z_{\rm init}=0.3$--0.5 Z$_\odot$, and SMC, $Z_{\rm init}=0.1$--0.2 Z$_\odot$
\cite{VanloonEtal99,Vanloon06} and other metal-poor dwarf galaxies
\cite{JacksonEtal07a,JacksonEtal07a,LagadecEtal07b}, and even if mass-loss
rates were reduced by an order of magnitude\footnote{Note that this is {\it
not} the case for red supergiants, which therefore explode as core-collapse
supernovae.}.

For low-mass AGB stars the preceding first ascent red giant branch (RGB) phase
also involves substantial mass loss. This affects the mantle mass with which
the star ascends the AGB, and thus its AGB evolution and mass loss. It has
been suggested recently \cite{Hansen05,KaliraiEtal07} that super-solar
metal-rich RGB stars may lose mass at a higher rate, truncating their RGB
evolution, evading the helium flash and producing low-mass white dwarfs. These
stars would never become AGB stars. Note that only a doubling of the mass lost
on the RGB is required for this to happen. RGB mass loss seems to be
fine-tuned within a narrow range though, as even in very metal-poor globular
clusters it is inferred to explain the blue horizontal branches whilst still
having to allow for the observed AGB stars and a PN in M\,15, $Z=0.005$
Z$_\odot$ (cf.\ \cite{VanloonEtal07}).

\section{What determines the mass-loss rate?}

Several mechanisms that could, in principle, drive mass loss do not work in
AGB stars. The main and efficient driver of hot star winds, radiation pressure
in atomic and ionic transitions does not work in AGB stars because they are
not hot enough to produce enough opacity and not luminous enough to provide
enough pressure. AGB stars do not reach the Eddington luminosity as defined in
terms of the radiation pressure in their optically thick mantles, nor do they
rotate fast enough to initiate or significantly facilitate mass loss.
Radiation pressure in molecular bands is not thought to be sufficient to
support a massive outflow either: because the bands are broad and the wind
will be slow, the wind never accelerates out of its own shadow, and therefore
only utilises the minor fraction of the luminosity that is emitted within the
molecular bands.

What does work, at least in some cases, is the two-stage mechanism of
pulsation-induced levitation, and radiation pressure on dust forming in this
cool, dense molecular atmosphere. The pulsation is the result of oscillations
in the partial ionization z\^one of hydrogen within the mantle, as the
ionization stage determines the opacity in the mantle and thus the balance
between gravity and the outward-directed radiation pressure, the so-called
$\kappa$-mechanism. The pulsation is stronger at lower temperature and higher
luminosity. Dust forms preferentially in the dense shocks during part of this
pulsation cycle (which has a timescale of $10^{2-3}$ days), and {\it
contributes} to driving a wind via radiation pressure and collisional coupling
with the gas \cite{BowenWillson91,WachterEtal02}.

The dust-driven wind regimes in the Hertzsprung-Russell diagram are set by the
Eddington luminosity above which the radiation pressure is sufficient to
support such a dust-driven wind \cite{FerrarottiGail06}. This depends on the
dust optical depth compared to the total mass in the wind, i.e.\ the opacity
of the dust species, dust:gas mass ratio and total mass-loss rate. Dust only
forms below a certain temperature and provided the density is high enough. On
the RGB, a dust-driven wind is only possible if carbonaceous dust is formed as
it has a relatively high opacity, but no carbon stars are found on the RGB
(although carbon stars do exist in old globular clusters,
\cite{VanloonEtal07}). Silicate dust can be used to drive a wind only for AGB
stars significantly more luminous than the tip of the RGB
\cite{FerrarottiGail06}. This also shows that the exact evolution of a star
in terms of its luminosity and especially temperature is critical for its
ability to form dust and use it to drive a wind.

What happens if the pulsation and/or dust formation fail? Because these tend
to be warmer, more compact stars, which are known to be chromospherically
active, it is believed that chromospheric processes are likely at the origin
of the driving mechanism. Instead of radiation, Alfv\'en (electromagnetic) or
acoustic (pressure) waves are invoked. Mass-loss rates are very difficult to
measure for these stars because of the sensitivity to the
excitation/ionization balance and deviations from local thermodynamic
equilibrium. A semi-empirical relation has been suggested by
\cite{SchroederCuntz05}, which would lead to higher mass-loss rates of
metal-poor stars, because they are warmer.

Observations of RGB stars do suggest that non-dusty winds may be at least as
efficient as dusty winds. Modelling of optical absorption line profiles in
globular cluster red giants by \cite{McdonaldVanloon07} yields mass-loss rates
in excess of Reimers' Law and in agreement with emission-wing diagnostics
\cite{Cohen76} and dust-derived mass-loss rates where available
\cite{OrigliaEtal02}. Indeed, \cite{JudgeStencel91} show that the combined
energy involved in radiation and wind is similar for chromospherically and
dust-driven winds, operating in different temperature regimes but with a
smooth transition z\^one (cf.\ \cite{Vanloon08}).

The luminosity and temperature are obvious measures of the radiation pressure
and ability to form dust. Because evolutionary tracks do not cross, the mass
is not needed as an independent parameter, $M=M(L,T)$. Both observations and
hydrodynamical computations support a parameterization of the mass-loss rate
of the form $\dot{M}(L,T)$, with a steep dependence on temperature but roughly
proportional to luminosity \cite{VanloonEtal99,VanloonEtal05,WachterEtal02}
--- where it must be realised that the temperature varies less than a factor
two.

The mass-loss rate has also been parameterized as a function of the pulsation
period, $P$ \cite{VassiliadisWood93}. This works reasonably well at least for
the sample for which it was derived, partly because $P=P(L,T)$ as it is a
measure of the stellar radius, $R$, and $L=4\pi \sigma R^2T^4$. The period
sets the timescale for dust to form and grow, the {\it amplitude} of pulsation
may be at least as relevant \cite{Vanloon08}.

\cite{JudgeStencel91} suggest that irrespective of the driving mechanism, the
mass-loss rate may be parameterized as a function of the surface gravity, $g$.
Again, it is not obvious whether the gravity, or escape velocity, is the
parameter of fundamental importance, as $g=g(L,T)$ because both the radius and
mass are functions of luminosity and temperature.

Complications arise if the AGB star is in a binary system; the mass-loss rate
may be enhanced equatorially, a circumbinary disc may trap ancient material,
a fast(er) bipolar wind may develop, and mass transfer may enrich the surface
of the companion star. In novae, mass loss from the donor indirectly
contributes to explosive nucleosynthesis, and a merger of the two stars may
induce an eruption.

\section{How does nucleosynthesis alter the conditions for the mass-loss
mechanism?}

Nucleosynthesis and thermal pulses affect the luminosity and temperature and
thus the mass-loss rate. The evolutionary track in a Hertzsprung-Russell
diagram depends on a star's initial metallicity, $Z$, helium abundance, and a
changing carbon:oxygen ratio. During a thermal pulse the luminosity and
temperature change in a complex manner, and depending on the $\dot{M}(L,T)$
formalism the star is normally expected to experience a brief episode of more
intense mass loss. \cite{VassiliadisWood93} predict less mass loss during the
first thermal pulses, but a rapid truncation of AGB evolution, compared to
Reimers' Law which allows more thermal pulses and a more balanced mass loss
over the AGB evolution. If a final thermal pulse occurs in a massive AGB star
after HBB ceased \cite{VanloonEtal98,FrostEtal98}, a massive carbon star is
formed and fresh dust will be carbonaceous. It is unclear whether this would
affect the mass-loss rate. A final thermal pulse occurring whilst on the
post-AGB track may sent the star back onto the AGB, rejuvenating its mass
loss. Clearly, thermal pulses could have a great impact if they occur in a
symbiotic system.

%
%
\begin{figure}[tb!]
\includegraphics[height=.505\textheight,clip=true]{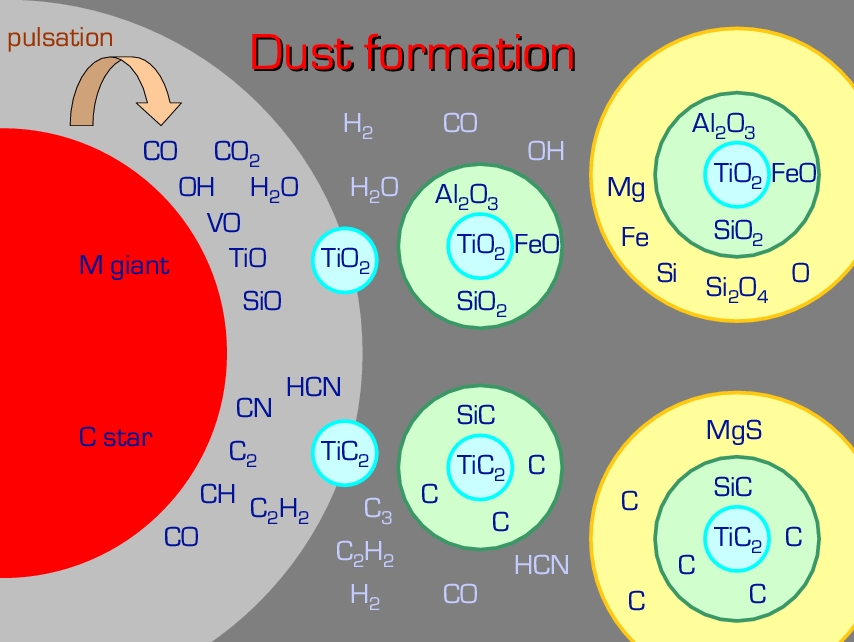}
\caption{Schematic overview of the dust formation process in oxygen- (top) and
carbon-dominated (bottom) circumstellar environments. Seeds involving trace
chemical elements such as titanium act as condensation nuclei. First coated
with highly refractory molecular species (e.g., alumina or silicon carbide),
these in turn are coated with more volatile species (e.g., silicates or
magnesium sulfide). The fraction of encapsulated ``proper'' metals such as
iron may vary, affecting the transparency of the grains. Little is known about
the initial nucleation and its relation to the composition of the molecular
atmosphere.}
\end{figure}

Dust formation is a complex process too. It takes place in an either
oxygen-dominated or carbon-dominated environment depending what element is
left after the other is exhausted in the formation of the carbon-monoxide
molecule (CO) in the cool, dense lower layers of the stellar atmosphere (Fig.\
1). Molecules do not directly bind together to form solid particles, but need
a nucleation seed. This is confirmed by meteoritic evidence (cf.\ Hoppe, these
proceedings) to be based on titanium, and possibly zirconium and silicon.
Neither of these are produced inside AGB stars, so their abundance is set by
the initial metallicity, $Z_{\rm init}$. The grain growth follows a sequence
of decreasing condensation temperature, with aluminium-oxides and
silicon-carbide (SiC) condensing before silicates and amorphous carbon (in
oxygen- and carbon-rich environments, respectively). Apart from composition
and size, the grain shape and degree of crystalline structure further affect
the optical properties and thus the ability of the grains to drive a wind.

%
%
\begin{figure}[tb!]
\includegraphics[height=0.365\textheight,clip=true]{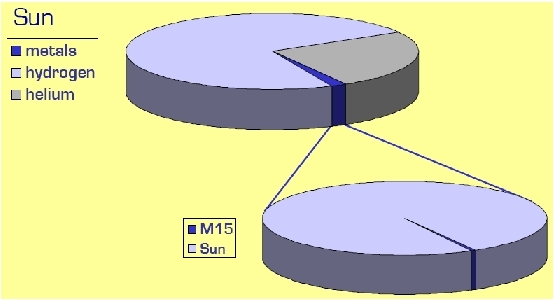}
\caption{Dust is formed out of metals, which constitute only a minor fraction
of any star. In metal-poor globular clusters, this fraction is truly minute
yet dust is observed in at least one such cluster, M\,15.}
\end{figure}

None of the dust species involve hydrogen or helium, which leaves it up to a
minor mass fraction of the wind to provide the agents for the interception of
the momentum from the radiation field. Even in the Sun it is only $\sim$1\%
(not all metals are used for dust formation, and CO takes away a large
fraction of it too), but in very metal-poor environments such as in M\,15,
with $Z=0.005$ Z$_\odot$, dust is seen \cite{BoyerEtal06}. To appreciate how
little mass in condensable material this is, just look at Fig.\ 2!

Oxygen-rich AGB stars pose a challenge: \cite{Woitke06} reminds us of earlier
findings that silicates are transparent; iron seems to be required to provide
sufficient opacity where the stellar radiation peaks. \cite{HoefnerAndersen07}
propose an alternative solution, if some carbonaceous dust could still be
formed. Although most of the dust mass would be oxygenous, most of the opacity
in the near-infrared would be due to carbonaceous dust.

Oxygenous minerals present a rich diversity, due to various admixtures of
aluminium, silicon, iron, magnesium, even calcium. Mixing of the products of
CNO processing could affect the amount of oxygen available, but it is
generally the other elements it binds with, that are the critical component
(Fig.\ 3). Changes in the aluminium abundance could affect the initial coating
of the nucleation seeds, but it is unclear whether the later coating by
silicates is affected. Hence, titanium is critical as seed, silicon for grain
growth, and iron for opacity. All these scale directly with initial
metallicity. This is consistent with the simple and predicted behaviour of
dust-driven winds of massive AGB stars: $\dot{M}$ is insensitive to $Z_{\rm
init}$, but the dust:gas ratio $\psi\propto Z_{\rm init}$
\cite{Vanloon00,Vanloon06,MarshallEtal04}.

%
%
\begin{figure}[tb!]
\includegraphics[height=.349\textheight,clip=true]{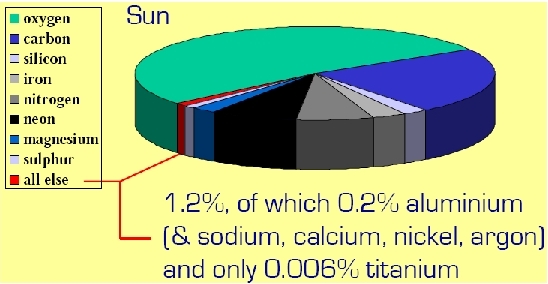}
\caption{Approximate fractions of the metals in the Sun, of various elements.
Oxygen and/or carbon, which form the basis of most of the circumstellar dust
produced in AGB winds, are always abundant. Important constituents of certain
dust species, iron, silicon, magnesium and aluminium are much less abundant
and thus limit grain growth. As a nucleation seed, titanium is a key element
but extremely rare.}
\end{figure}

Metal-poor carbon stars exhibit strong molecular absorption
\cite{VanloonZijlstraGroenewegen99,MatsuuraEtal05,MatsuuraEtal06,VanloonEtal06},
but there is no convincing evidence for enhanced dust formation
\cite{SloanEtal06,ZijlstraEtal06,LagadecEtal07b,MatsuuraEtal07}. These works
show that, in sufficiently cold envelopes, magnesium-sulfide (MgS) forms in a
similar ratio to the amorphous carbon which constitutes the bulk of
carbonaceous dust. This suggests that if MgS is depleted in metal-poor winds
because of lower abundances of magnesium and sulphur, amorphous carbon must be
depleted too. SiC is more complicated to interpret as it may be coated by
amorphous carbon.

In carbon star winds too, the nucleation seeds depend on titanium and similar
elements. The strong molecular bands especially of the slightly larger
molecules such as acetylene and C$_3$ suggest that the density of molecules
may not be larger, but they may be found throughout a larger part of the wind
\cite{VanloonEtal06}. The grain growth may thus not be any different. Indeed,
the observed spectral energy distributions are consistent with $\psi\propto
Z_{\rm init}$ and $\dot{M}$ insensitive to $Z_{\rm init}$
\cite{Vanloon00,Vanloon06}, and this is also supported by considerations of
the molecular mass-loss rate \cite{MatsuuraEtal06}, the availability of
silicon in gas phase and SiC \cite{MatsuuraEtal07}, and the lower dust
production seen in magellanic PNe \cite{StanghelliniEtal07}.

S stars, which have a carbon:oxygen ratio approaching unity, are a relatively
unimportant contributor to chemical enrichment and unimportant phase for the
star itself, because the mass-loss rate is not particularly high and the phase
lasts very few thermal pulse cycles. They may teach us something about dust
formation and growth, though, when oxygen or carbon have become critical
elements in the chain of events.

\begin{theacknowledgments}
I would like to thank Maurizio Busso for inviting me to the workshop to talk
about AGB mass loss, all the participants for a very interesting and joyous
time together, and Joana for putting up admirably with my busy travel diary.
This talk was presented on 26 October 2007, which was also the ``Wear it
Pink'' awareness day of the british Breast Cancer Campaign, to which this talk
is dedicated. Oh yes, and the food was very good!
\end{theacknowledgments}

\bibliographystyle{aipproc}                           

\end{document}